\documentclass[journal,12pt,twoside,onecolumn]{IEEEtran}
\usepackage{cite}
\ifCLASSINFOpdf
  \usepackage[pdftex]{graphicx}
  \DeclareGraphicsExtensions{.pdf,.jpeg,.png}
\else
  \usepackage[dvips]{graphicx}
  \DeclareGraphicsExtensions{.eps,.ps}
\fi
\usepackage[cmex10]{amsmath}
\usepackage{amsfonts}
\usepackage{amsthm}
\usepackage{array}
\usepackage[tight,footnotesize]{subfigure}
\usepackage{url}
\usepackage{verbatim}

\newtheorem{prop}{Proposition}
\newtheorem{lemma}{Lemma}

\theoremstyle{remark}
\newtheorem*{remark}{Remark}
\DeclareMathOperator{\E}{\mathbb{E}}

\DeclareMathOperator*{\argmin}{\arg\,\min}

\DeclareMathOperator{\var}{var}
\DeclareMathOperator{\mbbP}{\mathbb{P}}

\newcommand{\mbmu}{\boldsymbol{\mu}}

\newcommand{\mbv}{\mathbf{v}}

\newcommand{\mby}{\mathbf{y}}

\newcommand{\mbu}{\mathbf{u}}

\newcommand{\mbp}{\mathbf{p}}

\newcommand{\mbY}{\mathbf{Y}}

\newcommand{\mbsigma}{\boldsymbol{\sigma}}

\newcommand{\mbxi}{\boldsymbol{\xi}}

\newcommand{\Hh}{\hat{H}}

\newcommand{\ulu}{\underline{u}}
\newcommand{\olu}{\overline{u}}
\newcommand{\oly}{\overline{y}}

\newcommand{\mcY}{\mathcal{Y}}

\newcommand{\mcN}{\mathcal{N}}

\begin{document}
%
% paper title
% can use linebreaks \\ within to get better formatting as desired
\title{Adaptive Sensing Resource Allocation Over Multiple Hypothesis Tests}
%
%
% author names and IEEE memberships
% note positions of commas and nonbreaking spaces ( ~ ) LaTeX will not break
% a structure at a ~ so this keeps an author's name from being broken across
% two lines.
% use \thanks{} to gain access to the first footnote area
% a separate \thanks must be used for each paragraph as LaTeX2e's \thanks
% was not built to handle multiple paragraphs
%

%\author{Dennis~Wei~and~Alfred~O.~Hero,~III}
\author{Dennis~Wei%,~\IEEEmembership{Member,~IEEE,}
%        and~Alfred~O.~Hero,~III%,~\IEEEmembership{Fellow,~IEEE}% <-this % stops a space
%\thanks{%Manuscript received \today. 
%This work was partially supported by Army Research Office grant W911NF-11-1-0391. 
\thanks{The author is with the Thomas~J.~Watson Research Center, IBM Research, Yorktown Heights, NY 10598, USA, e-mail: dwei@us.ibm.com.}}% <-this % stops a space
\maketitle

\begin{abstract}
%\boldmath
This paper considers multiple binary hypothesis tests with adaptive allocation of sensing resources from a shared budget over a small number of stages.  A Bayesian formulation is provided %that does not assume that the number of either true null or true alternative hypotheses is small.  
for the multistage allocation problem of minimizing the sum of Bayes risks, which is then recast as a dynamic program.  In the single-stage case, the problem is a non-convex optimization, for which an algorithm composed of a series of parallel one-dimensional minimizations is presented.  This algorithm ensures a global minimum %globally optimal solution  
under a sufficient condition.  In the multistage case, the approximate dynamic programming method of open-loop feedback control is employed.  In numerical simulations, the proposed allocation policies outperform alternative adaptive procedures when the numbers of true null and alternative hypotheses are not too imbalanced.  In the case of few alternative hypotheses, the proposed policies are competitive using only a few stages of adaptation.  In all cases substantial gains over non-adaptive sensing are observed. 
\end{abstract}
% IEEEtran.cls defaults to using nonbold math in the Abstract.
% This preserves the distinction between vectors and scalars. However,
% if the journal you are submitting to favors bold math in the abstract,
% then you can use LaTeX's standard command \boldmath at the very start
% of the abstract to achieve this. Many IEEE journals frown on math
% in the abstract anyway.

% Note that keywords are not normally used for peerreview papers.

% For peer review papers, you can put extra information on the cover
% page as needed:
% \ifCLASSOPTIONpeerreview
% \begin{center} \bfseries EDICS Category: 3-BBND \end{center}
% \fi
%
% For peerreview papers, this IEEEtran command inserts a page break and
% creates the second title. It will be ignored for other modes.
\IEEEpeerreviewmaketitle

% needed in second column of first page if using \IEEEpubid
%\IEEEpubidadjcol

\section{Introduction}
\label{sec:intro}

This paper is concerned with the problem of multiple binary hypothesis tests under a shared sensing budget.  Sensing resources can be allocated adaptively over multiple stages to the hypothesis tests, taking past observations into account.  Intuitively, the advantage of %such 
adaptive allocation is that resources can be continually shifted from tests where the outcome is more certain to those that are less certain.  For example, in wide-area search and surveillance, sensors can be directed to gradually concentrate more time, samples, or energy on spatial regions where target presence is the most uncertain.  Other applications include adaptive spectrum sensing for unoccupied communication bands \cite{tajer2012}, biomedical clinical trials with multiple endpoints \cite{jennison2000}, and multistage gene association studies \cite{zehetmayer2008}. 

Adaptive and sequential methods for multiple hypothesis testing have been studied recently in \cite{haupt2011,malloy2011b,malloyarXiv2012,bartroff2010,de2012,bartroffJSPI2014,bartroff2013,bartroffarXiv2014}.  One set of papers \cite{haupt2011,malloy2011b,malloyarXiv2012} %particularly in the context of 
focuses on support recovery for sparse signals.  These works showed that simple multistage thresholding procedures can asymptotically drive error rates to zero with slower growth in resources compared to non-adaptive procedures; \cite{haupt2011} focused on Gaussian observations and false discovery/non-discovery rates (FDR/FNR), while \cite{malloy2011b,malloyarXiv2012} considered more general likelihoods %model  
and the family-wise error rate (FWER).  The present work differs from and adds to \cite{haupt2011,malloy2011b,malloyarXiv2012} in three major respects:  First, no sparsity assumption is made on the number of alternative (or null) hypotheses that are true.  Indeed, significant performance gains %over non-adaptive sensing 
are demonstrated even when the hypotheses occur in equal numbers.  Second, the number of stages, i.e., the number of opportunities to adapt, is decoupled from the number of hypothesis tests and is deliberately constrained to be small.  It is shown that much of the benefit of adaptation can be realized with only two or three stages.  Third, a Bayesian formulation is adopted %herein 
that allows for composite null and alternative hypotheses %both null and alternative hypotheses to be composite 
given %some 
statistical prior knowledge; \cite{haupt2011,malloy2011b,malloyarXiv2012} in contrast require a simple null hypothesis but less prior information.

A second series of works \cite{bartroff2010,de2012,bartroffJSPI2014,bartroff2013,bartroffarXiv2014} has developed sequential tests %of multiple hypotheses 
that control various multiple testing error metrics: FWER \cite{bartroff2010}, both type I and type II FWER simultaneously \cite{de2012,bartroffJSPI2014}, FDR and FNR simultaneously \cite{bartroff2013}, and $k$-FWER or $\gamma$-FDP \cite{bartroffarXiv2014}.  These procedures permit general likelihoods and dependences between the multiple data sequences, leveraging existing methods to control %Type I and II 
sequential error rates for individual data sequences on the one hand, and the multiple testing error rates mentioned above on the other hand.  Unlike \cite{haupt2011,malloy2011b,malloyarXiv2012} and this work, \cite{bartroff2010,de2012,bartroffJSPI2014,bartroff2013,bartroffarXiv2014} focus on sequential procedures, which allow an indefinite number of stages at which sensing decisions can be made, subject to ensuring (perhaps conservatively) that the desired error rates are below specified levels.  In contrast, in \cite{haupt2011,malloy2011b,malloyarXiv2012} and herein, both the number of stages and the resource budget are fixed while the error rates are minimized.  This non-sequential setting also gives rise to the problem of resource allocation over stages and tests, which is not considered in \cite{bartroff2010,de2012,bartroffJSPI2014,bartroff2013,bartroffarXiv2014}.

The present paper and \cite{haupt2011,malloy2011b,malloyarXiv2012,bartroff2010,de2012,bartroffJSPI2014,bartroff2013,bartroffarXiv2014} are related more broadly %draw inspiration from classical work in 
to the literature on %sequential (single) hypothesis testing 
(single/non-multiple) sequential tests \cite{wald1948}, especially %tests 
with more than two hypotheses and control over observations \cite{chernoff1959,bessler1960,naghshvarAnnStat2013,naghshvarJSTSP2013,nitinawarat2013}.  However, while it may be possible in principle to apply these methods for more than two hypotheses to the multiple testing problem, performance losses may be expected compared to more specialized methods such as in \cite{haupt2011,malloy2011b,malloyarXiv2012,bartroff2010,de2012,bartroffJSPI2014,bartroff2013,bartroffarXiv2014} and herein.  Moreover, %a distinction should be drawn between 
%sequential procedures as defined in 
the procedures in \cite{chernoff1959,bessler1960,naghshvarAnnStat2013,naghshvarJSTSP2013,nitinawarat2013} are sequential in the sense of the previous paragraph, %, which 
%allow an indefinite number of stages at which sensing decisions can be made, %and 
again in contrast to the non-sequential 
approach in this paper. 
%where the number of stages is fixed and small and the resource budget is also fixed.  
In addition, \cite{chernoff1959,bessler1960,naghshvarAnnStat2013,naghshvarJSTSP2013,nitinawarat2013} consider a finite number of sensing choices of differing quality but equal cost, whereas in this work the sensing control is continuous-valued and observation quality is a direct function of resource cost.

The statistical model and dynamic programming methods in this paper are similar to those in \cite{adapEstJSTSP2013} (except for the sparsity assumption).  However, the objective of hypothesis testing differs significantly from \cite{adapEstJSTSP2013}, which focuses on amplitude estimation of sparse signals.  This difference has an important consequence for optimization: the Bayes risk adopted here as the %testing 
performance metric is not a convex function of the resource allocations, unlike the estimation error metrics in \cite{adapEstJSTSP2013}.  The lack of convexity complicates the resource allocation problem and necessitates %, thus making the development of 
an alternative optimization method. %a major focus of the current paper.

Section \ref{sec:prob} presents a Bayesian formulation of multiple binary hypothesis testing with adaptive allocation of sensing resources from a fixed budget.  Only Gaussian observations are considered in this paper.  The multistage allocation problem of minimizing the sum of Bayes risks is then recast as a dynamic program.  In Section \ref{sec:policy}, single-stage and multistage solutions %to the dynamic program 
are developed.  In the single-stage case, an algorithm is proposed involving %multiple 
parallel single-variable minimizations and an outer search over a %single 
Lagrange multiplier.  Despite the non-convexity of the Bayes risk objective function as noted earlier, this algorithm can %nevertheless 
guarantee a global minimum when a sufficient condition is met.  In the multistage case, a tractable approximate solution is proposed using open-loop feedback control \cite{bertsekas2005} with the property of monotonic improvement as the number of stages increases, similar to \cite{adapEstJSTSP2013}.  Section~\ref{sec:num} presents numerical simulations comparing the proposed allocation policies to \cite{haupt2011,malloyarXiv2012}, demonstrating advantages %.   The results demonstrate the advantage of the proposed approach 
when the numbers of null and alternative hypotheses are within an order of magnitude of each other.  In the highly imbalanced case, the proposed policies remain competitive and achieve most of the gains using two or three stages.

\section{Problem formulation}
\label{sec:prob}

We consider $n$ binary hypothesis tests indexed by $i = 1,\dots,n$.  A priori, the $i$th null and alternative hypotheses are true with known probabilities $\mbbP(H_{i} = 0) = 1 - p_{i}(0)$ and $\mbbP(H_{i} = 1) = p_{i}(0)$, and $H_{i}$, $H_{j}$ are statistically independent for $i \neq j$.  It is not assumed that $p_{i}(0) \ll 1$, i.e., the alternative hypothesis is not necessarily rare, unlike in \cite{haupt2011,malloy2011b,malloyarXiv2012}.  

Observations are made in $T$ stages (indexed in parentheses) following a model similar to the one in \cite{adapEstJSTSP2013}. %The stage index is denoted in parentheses.  
The quality of each observation is controlled by the amount of sensing resources allocated to it.  Specifically, given resource $u_{i}(t-1) > 0$, the observation $y_{i}(t)$ for test $i$ in stage $t$ is conditionally distributed as 
\begin{equation}\label{eqn:obs}
y_{i}(t) \mid x_{i}, u_{i}(t-1) \sim \mcN(x_{i}, \nu^{2} / u_{i}(t-1)), \quad t = 1,\dots,T,
\end{equation}
so that the precision (inverse variance) increases with $u_{i}(t-1)$.  If $u_{i}(t-1) = 0$, the observation $y_{i}(t)$ is not taken.  The mean $x_{i}$ depends on $H_{i}$ as specified in \eqref{eqn:x} below.  The nominal variance $\nu^{2}$ is assumed to be known.  The observations $y_{i}(t)$ are independent across tests $i$ and conditionally independent across stages $t$ given $x_{i}$ and $u_{i}(t)$, $t = 0,\dots,T-1$ (but not unconditionally independent).

As an example of the observation model above with $u_{i}(t-1)$ an integer, \eqref{eqn:obs} results if $u_{i}(t-1)$ i.i.d.\ observations, each distributed as $\mcN(x_{i}, \nu^{2})$, are taken in stage $t$ and $y_{i}(t)$ is computed as the sample mean (a sufficient statistic for $x_{i}$).  More generally, $u_{i}(t-1)$ is allowed to take on any non-negative real value to model continuous-valued resources and for mathematical convenience.  The resource allocations are constrained by an overall deterministic budget,
\begin{equation}\label{eqn:budget}
\sum_{t=0}^{T-1} \sum_{i=1}^{n} u_{i}(t) = Bn,
\end{equation}
so that the average budget per test is $B$.  This budget constraint couples the hypothesis tests together. 

In adaptive sensing, resource allocations can depend causally on all previous observations.  Define $\mby(t) = (y_{1}(t), \dots, y_{n}(t))$ (similarly for other vectors) and $\mbY(t) = \{\mby(1),\dots,\mby(t)\}$. %we have that 
Then $u_{i}(t-1)$ in \eqref{eqn:obs} is in general a function of $\mbY(t-1)$.  The mappings $\mbY(t) \mapsto \mbu(t)$ are referred to as the resource allocation policy.  

The mean parameters $x_{i}$ in \eqref{eqn:obs} are independent over $i$ and follow Gaussian distributions conditioned on $H_{i}$,
\begin{equation}\label{eqn:x}
x_{i} \mid H_{i} \sim \mcN\bigl( \mu_{i}^{H_{i}}(0), \sigma_{i}^{H_{i}}(0)^{2} \bigr), \quad H_{i} = 0, 1,
\end{equation}
with known prior parameters $\mu_{i}^{H_{i}}(0)$ and $\sigma_{i}^{H_{i}}(0)^{2}$.  Hence both the null and alternative hypotheses can be composite if $\sigma_{i}^{0}(0), \sigma_{i}^{1}(0) > 0$, generalizing \cite{adapEstJSTSP2013}.  By %translating and 
interchanging if necessary, it is assumed that %can simplify notation by taking $\mu_{i}^{0}(0) = 0$, $\mu_{i}(0) = \mu_{i}^{1}(0)$ and assuming 
$\sigma_{i}^{0}(0) \leq \sigma_{i}^{1}(0)$ without loss of generality. 

After all observations have been collected, a decision $\Hh_{i}(T) : \mbY(T) \mapsto \{0,1\}$ is made in each of the  hypothesis tests.  Performance is measured by the sum of Bayes risks,
\begin{equation}\label{eqn:BayesRisk}
R = \sum_{i=1}^{n} \E_{H_{i}, \mbY(T)}\left[ %(1-p_{i}(0)) \mbbP(\Hh_{i}=1 \mid H_{i}=0) + c p_{i}(0) \mbbP(\Hh_{i}=0 \mid H_{i}=1) 
(1 - H_{i}) \Hh_{i}(T) + c H_{i} \bigl( 1 - \Hh_{i}(T) \bigr)
\right],
\end{equation}
where $\E_{H_{i}, \mbY(T)}$ denotes expectation over $H_{i}$ and $\mbY(T)$, and $c$ is the cost of a Type II error (miss) relative to a Type I error (false alarm).  For $c = 1$, \eqref{eqn:BayesRisk} is the sum of the probabilities of error in each test, which is a union bound on the family-wise error rate, i.e., the probability of any error.  It is also possible to minimize the family-wise error rate directly using an approach similar to the one herein, but this is not developed further. 

In summary, the problem is to minimize the Bayes risk sum \eqref{eqn:BayesRisk} with respect to the resource allocation policy $\{\mbu(t)\}$ subject to the total budget constraint \eqref{eqn:budget}. 

\subsection{Dynamic programming formulation}
\label{subsec:probDP}

Similar to \cite{adapEstJSTSP2013}, the multistage minimization of the Bayes risk sum $R$ can be cast as a dynamic program \cite{bertsekas2005}, where the state is a belief state summarizing the posterior distributions of $H_{i}$ and $x_{i}$ given observations $\mbY(t)$.  Using \cite[Lem.~1]{adapEstJSTSP2013} to derive these posterior distributions, it can be shown that the variables $H_{i} \mid \mbY(t)$ remain independent over $i$ with parameters $p_{i}(t) = \mbbP(H_{i} = 1 \mid \mbY(t))$, and $x_{i} \mid H_{i}, \mbY(t)$ remain independent Gaussian with means $\mu_{i}^{H_{i}}(t) = \E\left[ x_{i} \mid H_{i}, \mbY(t) \right]$ and variances $\sigma_{i}^{H_{i}}(t)^{2} = \var\left( x_{i} \mid H_{i}, \mbY(t) \right)$.  The posterior parameters evolve according to 
\begin{subequations}\label{eqn:postEvol}
\begin{align}
p_{i}(t+1) &= \frac{p_{i}(t) f_{i}^{1}\bigl( y_{i}(t+1); t \bigr)}{p_{i}(t) f_{i}^{1}\bigl( y_{i}(t+1); t \bigr) + (1 - p_{i}(t)) f_{i}^{0}\bigl( y_{i}(t+1); t \bigr)},\label{eqn:pEvol}\\
\mu_{i}^{H_{i}}(t+1) &= \frac{\nu^{2} \mu_{i}^{H_{i}}(t) + \sigma_{i}^{H_{i}}(t)^{2} u_{i}(t) y_{i}(t+1)}{\nu^{2} + \sigma_{i}^{H_{i}}(t)^{2} u_{i}(t)},\\
\sigma_{i}^{H_{i}}(t+1)^{2} &= \frac{\nu^{2} \sigma_{i}^{H_{i}}(t)^{2}}{\nu^{2} + \sigma_{i}^{H_{i}}(t)^{2} u_{i}(t)},
\end{align}
\end{subequations}
where in \eqref{eqn:pEvol}, $f_{i}^{H_{i}}(\cdot; t)$ is the probability density function (PDF) of 
\begin{equation}\label{eqn:yPost}
y_{i}(t+1) \mid H_{i}, \mbY(t) \sim \mcN\bigl( \mu_{i}^{H_{i}}(t), \sigma_{i}^{H_{i}}(t)^{2} + \nu^{2} / u_{i}(t) \bigr).
\end{equation}
The index $t = 0$ corresponds to the prior parameters in effect before any observations are taken.  

Define the belief state as $\mbxi(t) = (\mbp(t), \mbmu(t), \mbsigma(t)^{2}, U(t))$, where $\mbmu(t)$ and $\mbsigma(t)^{2}$ include all components indexed by $i$ and $H_{i} = 0,1$, and $U(t)$ is the resource budget remaining in stage $t$ with $U(0) = Bn$.  %This state definition fulfills the requirement for a dynamic program as specified below. It is shown that $R$ in \eqref{eqn:BayesRisk} is a function of $\mbxi(T-1)$ and the control $\mbu(T-1)$ alone, thus satisfying the requirement for a dynamic program.  
This state definition makes the objective function additive over stages, as required for a dynamic program.  In fact the only explicit dependence is on the last stage, as specified below. 
\begin{prop}
The Bayes risk sum \eqref{eqn:BayesRisk} is the expected value of a function only of the state $\mbxi(T-1)$ and control $\mbu(T-1)$, %at the same stage index,
%\begin{multline}\label{eqn:BayesRiskDP}
\begin{equation}\label{eqn:BayesRiskDP}
R = \sum_{i=1}^{n} \E_{\mbY(T-1)} \left[ \int_{-\infty}^{\infty} \min\left\{ \bigl(1-p_{i}(T-1)\bigr) f_{i}^{0}(y; T-1), 
c p_{i}(T-1) f_{i}^{1}(y; T-1) \right\} \, dy \right],
%\end{multline}
\end{equation}
where the PDFs $f_{i}^{0}(\cdot; T-1)$ and $f_{i}^{1}(\cdot; T-1)$ are completely parameterized in \eqref{eqn:yPost} by $\mbxi(T-1)$ and $\mbu(T-1)$.
\end{prop}

\begin{IEEEproof}
Each of the Bayes risks in \eqref{eqn:BayesRisk} is minimized by the weighted maximum a posteriori (MAP) rule.
Using the definition of $p_{i}(T)$, the $i$th term in \eqref{eqn:BayesRisk} can thus be rewritten as %$\E_{\mbY(T)} \left[ \min\{ 1 - p_{i}(T), c p_{i}(T) \} \right]$. %the following expectation over $\mbY(T)$: 
\begin{equation}\label{eqn:BayesRiskMAP}
\E_{\mbY(T)} \left[ \min\{ 1 - p_{i}(T), c p_{i}(T) \} \right].
%\sum_{i=1}^{n} \E\left[ \min\{ 1 - p_{i}(T), c p_{i}(T) \} \right].
\end{equation}
Next we substitute for $p_{i}(T)$ using \eqref{eqn:pEvol} and iterate expectations over $y_{i}(T) \mid \mbY(T-1)$ and then $\mbY(T-1)$ to obtain 
\begin{multline}\label{eqn:BayesRiskDP1}
\E_{\mbY(T-1)} \left[ \int_{-\infty}^{\infty} \frac{\min\{ (1 - p_{i}(T-1)) f_{i}^{0}\bigl( y_{i}(T); T-1 \bigr), c p_{i}(T-1) f_{i}^{1}\bigl( y_{i}(T); T-1 \bigr) \}}{p_{i}(T-1) f_{i}^{1}\bigl( y_{i}(T); T-1 \bigr) + (1 - p_{i}(T-1)) f_{i}^{0}\bigl( y_{i}(T); T-1 \bigr)} \right.\\
\left. \phantom{\int} \times f\bigl( y_{i}(T) \mid \mbY(T-1) \bigr) dy_{i}(T) \right],
\end{multline}
where the inner expectation has been expressed as an explicit integral. The denominator in \eqref{eqn:BayesRiskDP1} can be recognized as the PDF of $y_{i}(T) \mid \mbY(T-1)$, yielding \eqref{eqn:BayesRiskDP} after cancellation. 
%\begin{multline}\label{eqn:BayesRiskDP}
%R = \sum_{i=1}^{n} \E_{\mbY(T-1)} \left[ \int_{-\infty}^{\infty} \min\left\{ \bigl(1-p_{i}(T-1)\bigr) f_{i}^{0}(y; T-1),\right.\right.\\
%\left.\phantom{\int}\bigl. c p_{i}(T-1) f_{i}^{1}(y; T-1) \bigr\} \, dy \right].
%\end{multline}
\end{IEEEproof}
\begin{remark}
The allocation $\mbu(T-1)$ is also constrained by the remaining budget $U(T-1)$, which is part of the state $\mbxi(T-1)$.  An equivalent unconstrained formulation can be obtained by augmenting \eqref{eqn:BayesRiskDP} with the stipulation that $R$ is infinite if $\sum_{i=1}^{n} u_{i}(T-1) > U(T-1)$, i.e., the budget is exceeded.
\end{remark}

\section{Resource allocation policies}
\label{sec:policy}

This section discusses single-stage and multistage resource allocation policies that minimize the Bayes risk sum \eqref{eqn:BayesRiskDP} under the budget constraint \eqref{eqn:budget}.  As discussed in Section%s~\ref{subsec:policy2} and 
~\ref{subsec:policyMulti}, the single-stage policy of Section~\ref{subsec:policy1} also applies to the last stage of any multistage policy.

\subsection{Single-stage policy}
\label{subsec:policy1}

In the single-stage case $T=1$, the expectation in \eqref{eqn:BayesRiskDP} is absent and the objective function simplifies.  The remaining integral is the Bayes risk of the optimal test between two Gaussian distributions with different means and variances.  %The Bayes risk can be evaluated by solving a quadratic inequality to determine the decision regions corresponding to the two terms in the minimization in \eqref{eqn:BayesRiskDP}, and then computing the Gaussian integrals, i.e., the Type I and Type II error probabilities.  These calculations are fairly standard and the details are omitted.  Here the integral in \eqref{eqn:BayesRiskDP} is simply denoted as 
Let $R_{i}(u_{i}; \mbxi_{i})$ denote this Bayes risk, where the stage index $T-1$ is suppressed to simplify notation, and $\mbxi_{i}$ represents the components of the state with index $i$.  Appendix~\ref{app:BayesRisk} provides explicit expressions for $R_{i}(u_{i}; \mbxi_{i})$ in terms of the standard Gaussian cumulative distribution function (CDF). The single-stage resource allocation problem is therefore 

\begin{equation}\label{eqn:singleStage}
R^{\ast}(\mbxi) = \min_{\mbu} \; \sum_{i=1}^{n} R_{i}( u_{i}; \mbxi_{i} ) \quad \text{s.t.} \quad \sum_{i=1}^{n} u_{i} = U, \;\; u_{i} \geq 0 \;\; \forall\; i.
\end{equation}

\begin{figure}[ht]
\centerline{
\subfigure[]{\includegraphics[width=0.40\columnwidth]{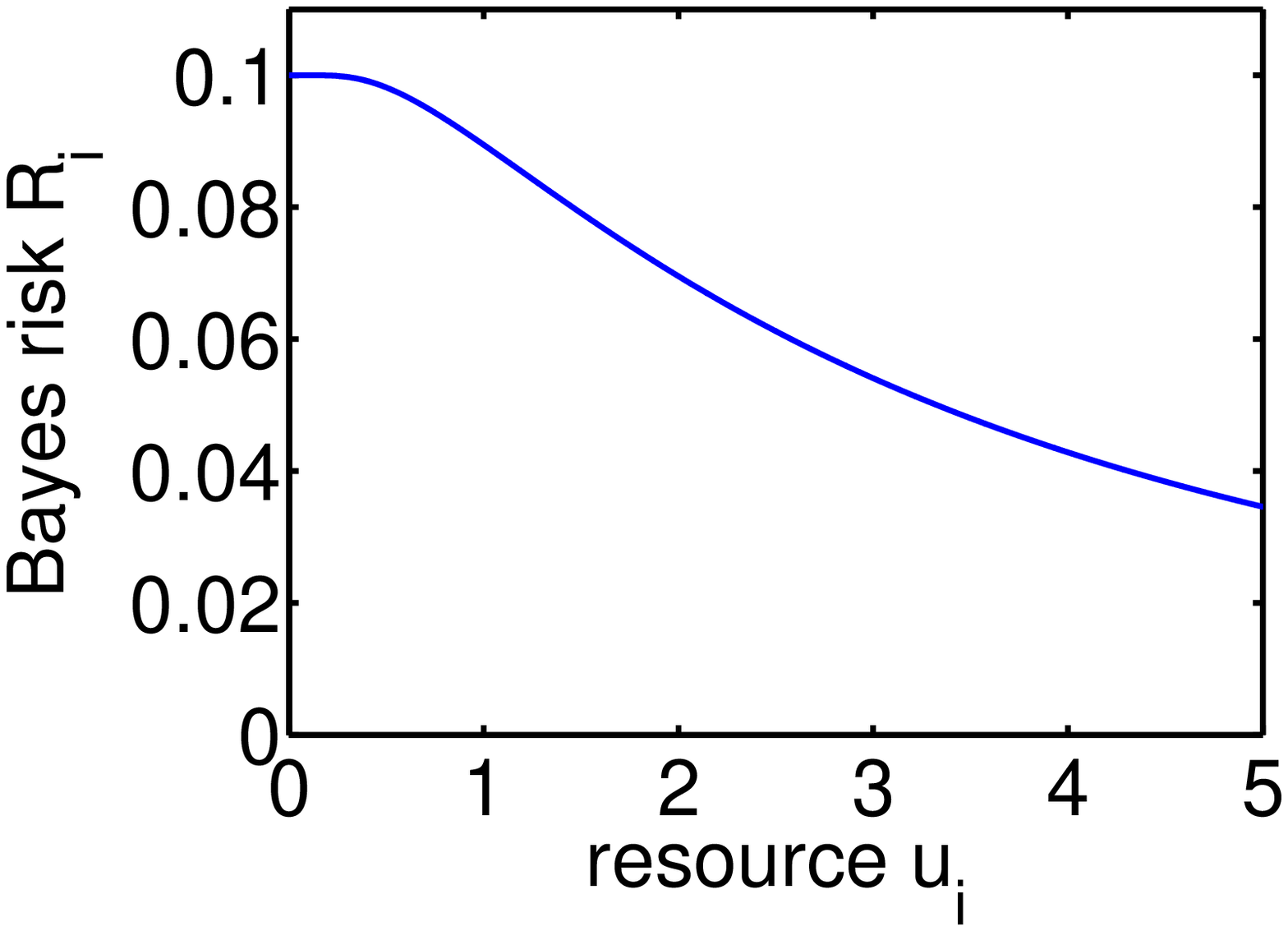}
\label{fig:Ri}}
\subfigure[]{\includegraphics[width=0.40\columnwidth]{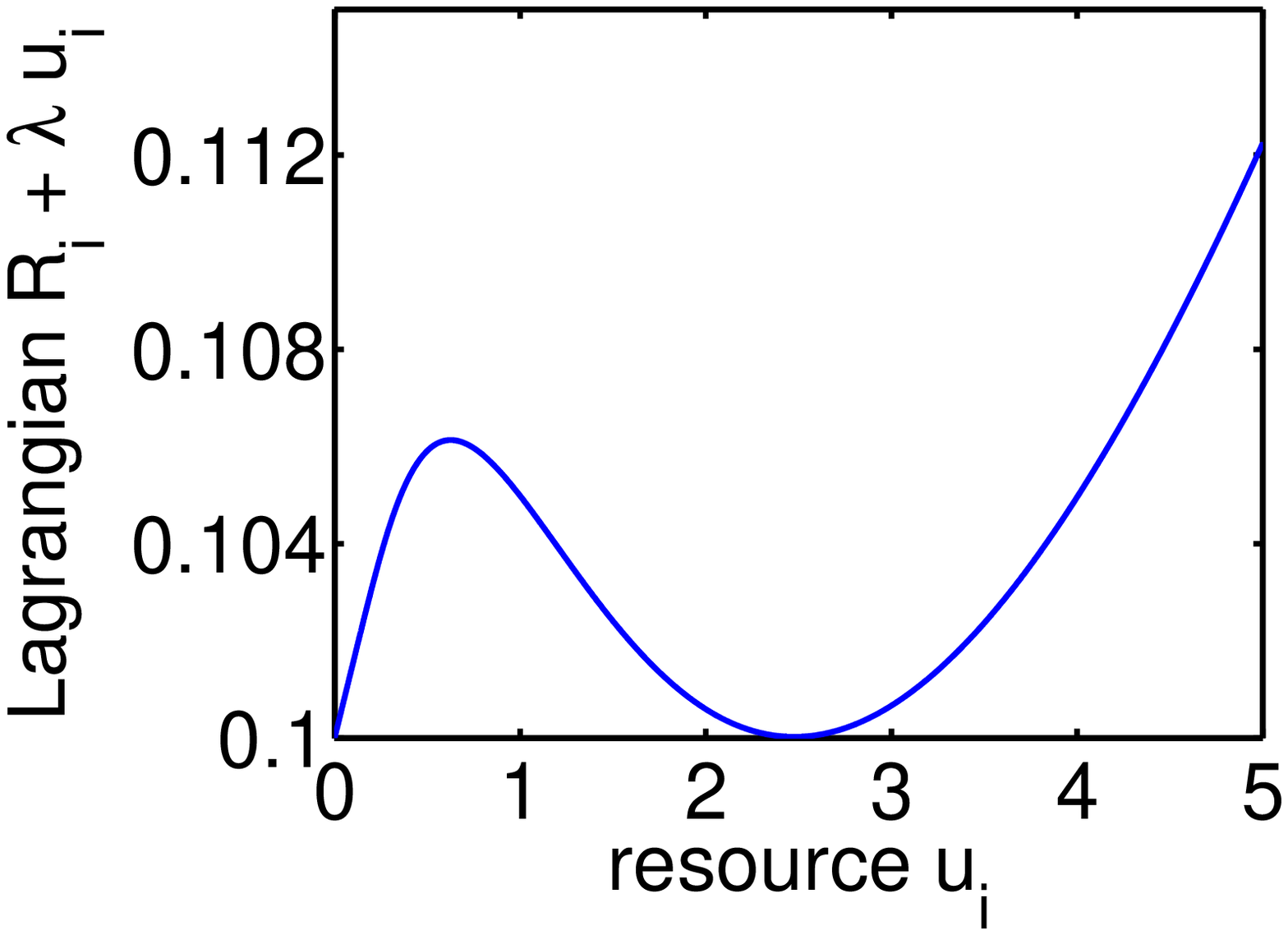}
\label{fig:RiL}}}
\caption{(a) The Bayes risk $R_{i}(u_{i}; \mbxi_{i})$ is a decreasing but non-convex function of $u_{i}$. (b) The Lagrangian in \eqref{eqn:RiL} can have more than one minimizer.}
\label{fig:RiRiL}
\end{figure}

Fig.~\ref{fig:Ri} shows that the Bayes risk $R_{i}( u_{i}; \mbxi_{i})$ is a decreasing but non-convex function of $u_{i}$ for a particular choice of parameters $\mbxi_{i}$.  These properties hold in general for other choices of $\mbxi_{i}$, implying that \eqref{eqn:singleStage} is a non-convex optimization problem.  

Despite the absence of convexity, it is still possible in some cases to guarantee a globally optimal solution to \eqref{eqn:singleStage}.  We consider minimizing a Lagrangian of \eqref{eqn:singleStage} in which only the equality constraint is dualized with Lagrange multiplier $\lambda$.  The Lagrangian then decouples over $i$.  Define the (possibly non-unique) minimizer of each Lagrangian component as 
\begin{equation}\label{eqn:RiL}
u_{i}(\lambda) \in \argmin_{u_{i} \geq 0} \; R_{i}(u_{i}; \mbxi_{i}) + \lambda u_{i}.
\end{equation}
Since $R_{i}(u_{i}; \mbxi_{i})$ is bounded from above by $\min\{1 - p_{i}, cp_{i}\} = R_{i}(0; \mbxi_{i})$, a negative value for $\lambda$ in \eqref{eqn:RiL} would result in divergence toward infinity.  Hence it is sufficient to consider $\lambda \geq 0$.  The following result gives a sufficient condition for $\mbu(\lambda) = \bigl(u_{1}(\lambda), \dots, u_{n}(\lambda)\bigr)$ to be globally optimal for \eqref{eqn:singleStage}.
\begin{prop}\label{prop:suffCond}
If there exists a Lagrange multiplier $\lambda \geq 0$ such that a set of minimizers $\mbu(\lambda) = \bigl(u_{1}(\lambda), \dots, u_{n}(\lambda)\bigr)$ defined by \eqref{eqn:RiL} is feasible for problem \eqref{eqn:singleStage}, then $\mbu(\lambda)$ is a global minimum of \eqref{eqn:singleStage}.
\end{prop}
\begin{IEEEproof}
This is an adaptation of \cite[Prop.~3.3.4]{bertsekas1999}, where the equality constraint $\sum_{i=1}^{n} u_{i} = U$ in \eqref{eqn:singleStage} (or equivalently two inequality constraints) has been incorporated into the Lagrangian function, while the remaining constraint set $X$ is the non-negative orthant.
\end{IEEEproof}
%Lemma~\ref{lem:suffCond} suggests that \eqref{eqn:singleStage} can be solved via a search over $\lambda$ and the minimizations in \eqref{eqn:RiL}.  To further specify the algorithm, the following monotonicity property and bound are used.
%\noindent 
The minimization in \eqref{eqn:RiL} also satisfies the monotonicity property below, which confirms the interpretation of $\lambda$ as a penalty parameter.
\begin{lemma}\label{lem:monotone}
If $\lambda_{1} < \lambda_{2}$, then $u_{i}(\lambda_{1}) \geq u_{i}(\lambda_{2})$ for any minimizers $u_{i}(\lambda_{1})$, $u_{i}(\lambda_{2})$ in \eqref{eqn:RiL}.
\end{lemma}
\begin{IEEEproof}
Let $u_{i}(\lambda_{1})$ be any minimizer in \eqref{eqn:RiL} for $\lambda = \lambda_{1}$.  Then for all $u_{i} > u_{i}(\lambda_{1})$, 
\begin{equation}\label{eqn:monotone1}
R_{i}\bigl( u_{i}(\lambda_{1}); \mbxi_{i} \bigr) + \lambda_{1} u_{i}(\lambda_{1}) \leq R_{i}\bigl( u_{i}; \mbxi_{i} \bigr) + \lambda_{1} u_{i}.
\end{equation}
By assumption, we have
\begin{equation}\label{eqn:monotone2}
(\lambda_{2} - \lambda_{1}) u_{i}(\lambda_{1}) < (\lambda_{2} - \lambda_{1}) u_{i}.
\end{equation}
Adding \eqref{eqn:monotone1} and \eqref{eqn:monotone2} yields 
\[
R_{i}\bigl( u_{i}(\lambda_{1}); \mbxi_{i} \bigr) + \lambda_{2} u_{i}(\lambda_{1}) < R_{i}\bigl( u_{i}; \mbxi_{i} \bigr) + \lambda_{2} u_{i},
\]
which implies that any minimizer $u_{i}(\lambda_{2})$ of \eqref{eqn:RiL} for $\lambda = \lambda_{2}$ must be no greater than $u_{i}(\lambda_{1})$.
\end{IEEEproof}

Based on Proposition~\ref{prop:suffCond} and Lemma~\ref{lem:monotone}, the following algorithm is proposed to solve \eqref{eqn:singleStage}, consisting of an outer bisection search over $\lambda$ and inner single-variable minimizations \eqref{eqn:RiL} to determine $u_{i}(\lambda)$, $i=1,\dots,n$, which can be done in parallel.  Lower and upper bounds $\ulu_{i}$ and $\olu_{i}$ are maintained on each $u_{i}$, where initially $\ulu_{i} = 0$ and $\olu_{i} = \infty$.  Any algorithm can be used to solve \eqref{eqn:RiL} subject to the bounds $\ulu_{i} \leq u_{i} \leq \olu_{i}$, for example gradient descent with logarithmically-spaced line search as used to generate the results in Section~\ref{sec:num}.  
Let $S(\lambda) = \sum_{i=1}^{n} u_{i}(\lambda)$.  If for a given $\lambda$, the resulting $u_{i}(\lambda)$ satisfy $S(\lambda) < U$, then $\lambda$ is decreased according to the bisection method, the lower bounds $\ulu_{i}$ are updated to the current solutions $u_{i}(\lambda)$, exploiting Lemma~\ref{lem:monotone}, and \eqref{eqn:RiL} is re-solved.  Analogous actions are taken if $S(\lambda) > U$.  If $S(\lambda) = U$, then by Proposition~\ref{prop:suffCond}, the algorithm terminates with a globally optimal solution to \eqref{eqn:singleStage}.

For the bisection search over $\lambda$, the initial lower bound is set at $0$.  The lemma below is used to set the initial upper bound.
\begin{lemma}\label{lem:UB}
Any minimizer $u_{i}(\lambda)$ in \eqref{eqn:RiL} is bounded from above as $u_{i}(\lambda) < R_{i}(0; \mbxi_{i}) / \lambda = \min\{ 1 - p_{i}, c p_{i} \} / \lambda$.
\end{lemma}
\begin{IEEEproof}
Since the Bayes risk $R_{i}(u_{i}; \mbxi_{i})$ is positive for finite $u_{i}$, if $u_{i} \geq R_{i}(0; \mbxi_{i}) / \lambda$ then $R_{i}(u_{i}; \mbxi_{i}) + \lambda u_{i} > R_{i}(0; \mbxi_{i})$ and $u_{i}$ cannot be minimal. 
\end{IEEEproof}
\noindent It follows that a sufficient upper bound on $\lambda$ is $\sum_{i=i}^{n} R_{i}(0; \mbxi_{i}) / U$, since any higher value can be seen to result in $S(\lambda) < U$.  Lemma~\ref{lem:UB} is also used to further constrain the inner minimizations over $u_{i}$ when it gives a tighter upper bound than $\olu_{i}$.

The above algorithm does not always ensure a global minimum for \eqref{eqn:singleStage}.  Specifically, it may not be possible to satisfy the condition in Proposition~\ref{prop:suffCond}, i.e., there is no $\lambda$ for which $S(\lambda) = U$ to make $\mbu(\lambda)$ feasible.  The problem is illustrated in Fig.~\ref{fig:RiL}, which shows a value for $\lambda$ such that the Lagrangian in \eqref{eqn:RiL} has two separated minimizers.  Any change in $\lambda$ would result in either the left or the right minimizer being unique.  Hence the function $S(\lambda)$ is discontinuous and the bisection search over $\lambda$ may not converge with $S(\lambda) = U$.  For the numerical results in Section~\ref{sec:num}, 
cases of non-convergence are addressed simply by rescaling the final solution $\mbu(\lambda)$ so that it sums to $U$.  The loss in optimality appears to be insignificant for large $n$ and can even be bounded analytically, although this is not presented here.

%\subsection{Two-stage policy}
%\label{subsec:policy2}

\subsection{Multistage policies}
\label{subsec:policyMulti}

In a multistage adaptive policy, the last-stage allocation $\mbu(T-1)$ can depend on all previous observations $\mbY(T-1)$.  In other words, $\mbu(T-1)$ is determined after conditioning on $\mbY(T-1)$, which again removes the expectation from \eqref{eqn:BayesRiskDP}.  Therefore the last-stage allocation problem in any multistage policy reduces to the single-stage case \eqref{eqn:singleStage}.  

For a two-stage policy, it remains to determine the first-stage allocation $\mbu(0)$.  This is done recursively by solving
\begin{equation}\label{eqn:firstStage}
\min_{\mbu(0)} \; \E_{\mby(1)} \left[ R^{\ast}(\mbxi(1)) \mid \mbxi(0), \mbu(0) \right] \quad \text{s.t.} \quad \sum_{i=1}^{n} u_{i}(0) \leq U(0), \;\; u_{i}(0) \geq 0 \;\; \forall\; i,
\end{equation}
where $R^{\ast}(\mbxi(1))$ is defined by \eqref{eqn:singleStage} as the optimal cost of the second stage, and the conditional notation reflects the parameterization of the distribution of $\mby(1)$ in terms of $\mbxi(0)$ and $\mbu(0)$ (see \eqref{eqn:yPost}).  In the case of priors that are homogeneous over $i$, i.e., $p_{i}(0)$, $\mu_{i}^{H_{i}}(0)$, $\sigma_{i}^{H_{i}}(0)^{2}$ do not depend on $i$ (but can depend on $H_{i}$), then the first-stage allocation is also homogeneous by symmetry, $u_{i}(0) = u(0)$, 
and \eqref{eqn:firstStage} becomes a scalar minimization with respect to $u(0) \in [0, U(0)/n]$.  This minimization is performed offline using Monte Carlo samples of $\mby(1)$ to approximate the expectation in \eqref{eqn:firstStage} and the algorithm in Section~\ref{subsec:policy1} to approximate $R^{\ast}(\mbxi(1))$ for each realization of $\mby(1)$.

For an inhomogeneous prior or more than two stages, an open-loop feedback control (OLFC) policy \cite{bertsekas2005} is employed, similar to \cite{adapEstJSTSP2013}.  %Although the derivation is slightly more involved than in \cite{adapEstJSTSP2013},% and is presented in \cite{}, 
%the end result is analogous.  
Consider the problem of determining the allocation $\mbu(t)$ in stage $t < T-1$ conditioned on available observations $\mbY(t)$ through the state $\mbxi(t)$. In exact dynamic programming, $\mbu(t)$ is optimized assuming that future allocations $\mbu(t+1),\dots,\mbu(T-1)$ are also chosen optimally as functions of $\mbY(t+1),\dots,\mbY(T-1)$ respectively.  However, in stage $t$ these future observations are not available and are therefore random quantities, which greatly complicates the optimization.  The OLFC simplification is to assume that $\mbu(t+1),\dots,\mbu(T-1)$ can depend only on current observations $\mbY(t)$, i.e., future planning is done ``open-loop''.  This leads to a joint optimization over $\mbu(t),\mbu(t+1),\dots,\mbu(T-1)$ of the Bayes risk sum \eqref{eqn:BayesRisk} conditioned on $\mbxi(t)$:
\begin{equation}\label{eqn:OLFC}
\min_{\mbu(t),\dots,\mbu(T-1)} \;\; \sum_{i=1}^{n} \E \left[ \min\{ 1 - p_{i}(T), c p_{i}(T) \} \mid \mbxi(t) \right] \quad \text{s.t.} \quad \sum_{i=1}^{n} \sum_{\tau=t}^{T-1} u_{i}(\tau) = U(t), \;\; u_{i}(\tau) \geq 0 \;\; \forall\; i, \tau,
\end{equation}
where \eqref{eqn:BayesRiskMAP} has been substituted into the objective function.  Once \eqref{eqn:OLFC} is solved, only the first stage $\mbu(t)$ is applied to collect new observations $\mby(t+1)$ as in \eqref{eqn:obs} and update the state to $\mbxi(t+1)$ using \eqref{eqn:postEvol}.  Then \eqref{eqn:OLFC} is solved for $\mbu(t+1),\dots,\mbu(T-1)$ given $\mbxi(t+1)$ under the same OLFC assumption, and the process continues.

The OLFC optimization problem \eqref{eqn:OLFC} can be further simplified to an instance of the single-stage optimization \eqref{eqn:singleStage}.  This together with Appendix~\ref{app:BayesRisk} provides an explicit expression for the objective function in terms of Gaussian CDFs, i.e.\ without expectation operators, and also reduces the number of optimization variables from $n(T-t)$ to $n$.
\begin{lemma}\label{lem:OLFC}
Let 
\[
v_{i}(t) = \sum_{\tau=t}^{T-1} u_{i}(\tau).
\]
The OLFC optimization problem \eqref{eqn:OLFC} reduces to an instance of the single-stage optimization \eqref{eqn:singleStage} with $u_{i} = v_{i}(t)$, $\mbxi = \mbxi(t)$, and $U = U(t)$.
\end{lemma}
\begin{IEEEproof}
The first step is to relate the Bayes risk objective in \eqref{eqn:OLFC} %posterior probability $p_{i}(T)$ 
to the state $\mbxi(t)$ in stage $t$.  Recalling the definition $p_{i}(t) = \mbbP(H_{i} = 1 \mid \mbY(t))$, an application of Bayes rule similar to \eqref{eqn:pEvol} yields  
\begin{align*}
p_{i}(T) &= \frac{p_{i}(t) f\bigl( y_{i}(t+1),\dots,y_{i}(T) \mid H_{i}=1, \mbY(t) \bigr)}{f\bigl( y_{i}(t+1),\dots,y_{i}(T) \mid \mbY(t) \bigr)},\\
1-p_{i}(T) &= \frac{(1 - p_{i}(t)) f\bigl( y_{i}(t+1),\dots,y_{i}(T) \mid H_{i}=0, \mbY(t) \bigr)}{f\bigl( y_{i}(t+1),\dots,y_{i}(T) \mid \mbY(t) \bigr)}.
\end{align*}
Hence 
\begin{align}
\E \left[ \min\{ 1 - p_{i}(T), c p_{i}(T) \} \mid \mbxi(t) \right] 
&= \int \dots \int \min\left\{ \frac{(1 - p_{i}(t)) f\bigl( y_{i}(t+1),\dots,y_{i}(T) \mid H_{i}=0, \mbxi(t) \bigr)}{f\bigl( y_{i}(t+1),\dots,y_{i}(T) \mid \mbxi(t) \bigr)} , \right.\nonumber\\
&\qquad\qquad \left. \frac{c p_{i}(t) f\bigl( y_{i}(t+1),\dots,y_{i}(T) \mid H_{i}=1, \mbxi(t) \bigr)}{f\bigl( y_{i}(t+1),\dots,y_{i}(T) \mid \mbxi(t) \bigr)} \right\}\nonumber\\
&\qquad\qquad \times f\bigl( y_{i}(t+1),\dots,y_{i}(T) \mid \mbxi(t) \bigr) dy_{i}(t+1) \dots dy_{i}(T)\nonumber\\
&= \int \dots \int \min\left\{ (1 - p_{i}(t)) f\bigl( y_{i}(t+1),\dots,y_{i}(T) \mid H_{i}=0, \mbxi(t) \bigr), \right.\nonumber\\
&\qquad\qquad \left. c p_{i}(t) f\bigl( y_{i}(t+1),\dots,y_{i}(T) \mid H_{i}=1, \mbxi(t) \bigr) \right\} dy_{i}(t+1) \dots dy_{i}(T).\label{eqn:BayesRiskOLFC}
\end{align}

To simplify \eqref{eqn:BayesRiskOLFC}, a Neyman factorization is derived for the joint density $f\bigl( y_{i}(t+1),\dots,y_{i}(T) \mid H_{i}, \mbxi(t) \bigr)$. Toward this end, we have 
\begin{equation}\label{eqn:yJoint}
f\bigl( y_{i}(t+1),\dots,y_{i}(T) \mid H_{i}, \mbxi(t) \bigr) = \int f\bigl( y_{i}(t+1),\dots,y_{i}(T) \mid x_{i}, \mbxi(t) \bigr) f\bigl( x_{i} \mid H_{i}, \mbxi(t) \bigr) dx_{i}.
\end{equation}
Under the OLFC assumption, conditioning on $\mbxi(t)$ also fixes the allocations $u_{i}(t),\dots,u_{i}(T-1)$.  Therefore $y_{i}(t+1),\dots,y_{i}(T) \mid x_{i}, \mbxi(t)$ are independent Gaussian according to \eqref{eqn:obs}.  Furthermore, it is straightforward to show that the weighted average 
\[
\oly_{i}(t) \equiv \frac{\sum_{\tau=t}^{T-1} u_{i}(\tau) y_{i}(\tau+1)}{\sum_{\tau=t}^{T-1} u_{i}(\tau)}
\]
is distributed as
\[
\oly_{i}(t) \mid x_{i}, \mbxi(t) \sim \mcN\left( x_{i}, \frac{\nu^{2}}{\sum_{\tau=t}^{T-1} u_{i}(\tau)} \right) = \mcN\left( x_{i}, \frac{\nu^{2}}{v_{i}(t)} \right)
\]
and is a sufficient statistic for $x_{i}$.  It follows that \eqref{eqn:yJoint} can be rewritten as 
\begin{align}
f\bigl( y_{i}(t+1),\dots,y_{i}(T) \mid H_{i}, \mbxi(t) \bigr) &= f\bigl( y_{i}(t+1),\dots,y_{i}(T) \mid \oly_{i}(t), \mbxi(t) \bigr)\nonumber\\
&\qquad\qquad \times \int f\bigl( \oly_{i}(t) \mid x_{i}, \mbxi(t) \bigr) f\bigl( x_{i} \mid H_{i}, \mbxi(t) \bigr) dx_{i}\nonumber\\
&= f\bigl( y_{i}(t+1),\dots,y_{i}(T) \mid \oly_{i}(t), \mbxi(t) \bigr) f\bigl( \oly_{i}(t) \mid H_{i}, \mbxi(t) \bigr),\label{eqn:NeymanFactor}
\end{align}
where
\begin{equation}\label{eqn:ybarPost}
\oly_{i}(t) \mid H_{i}, \mbxi(t) \sim \mcN\bigl( \mu_{i}^{H_{i}}(t), \sigma_{i}^{H_{i}}(t)^{2} + \nu^{2} / v_{i}(t) \bigr)
\end{equation}
as a result of compounding, similar to \eqref{eqn:yPost}.

The final step is to substitute the factorization \eqref{eqn:NeymanFactor} into \eqref{eqn:BayesRiskOLFC}.  Upon doing so, it is seen that the common factor $f\bigl( y_{i}(t+1),\dots,y_{i}(T) \mid \oly_{i}(t), \mbxi(t) \bigr)$ integrates to $1$, leaving 
\begin{multline*}%\label{eqn:BayesRiskOLFC2}
\E \left[ \min\{ 1 - p_{i}(T), c p_{i}(T) \} \mid \mbxi(t) \right] = \int \min\left\{ (1 - p_{i}(t)) f\bigl( \oly_{i}(t) \mid H_{i}=0, \mbxi(t) \bigr),\right.\\
\left. c p_{i}(t) f\bigl( \oly_{i}(t) \mid H_{i}=1, \mbxi(t) \bigr) \right\} d\oly_{i}(t).
\end{multline*}
Comparing the above expression with $R_{i}(u_{i}; \mbxi_{i})$, defined as the integral in \eqref{eqn:BayesRiskDP}, and \eqref{eqn:ybarPost} with \eqref{eqn:yPost}, we conclude that
\[
\E \left[ \min\{ 1 - p_{i}(T), c p_{i}(T) \} \mid \mbxi(t) \right] = R_{i}\bigl( v_{i}(t); \mbxi_{i}(t) \bigr).
\]
Rewriting the constraint in \eqref{eqn:OLFC} in terms of $v_{i}(t)$ completes the proof.
\end{IEEEproof}

According to Lemma~\ref{lem:OLFC}, the OLFC allocation in stage $t$ can be determined by first solving the single-stage problem \eqref{eqn:singleStage} with appropriate parameters.  However, the resulting solution $\mbv^{\ast}(t)$ does not specify the allocations of the sums $v_{i}(t) = \sum_{\tau=t}^{T-1} u_{i}(\tau)$ over stages, in particular the first allocation $\mbu(t)$ used to make new observations.  For this purpose, the approach in \cite{adapEstJSTSP2013} is followed in which $\mbu(t)$ is constrained to be a scaled version of $\mbv^{\ast}(t)$: 
$\mbu(t) = \beta(t; T) \mbv^{\ast}(t)$ where $\beta(t; T) \in [0,1]$ and the second argument $T$ denotes the number of stages in the policy.  Setting $\beta(t; T) < 1$ thus conserves some of the resource budget for future stages.  

The multipliers $\beta(t; T)$ are determined recursively for $T = 1,2,\dots$ as follows.  For $t = T-1$, $\mbv^{\ast}(T-1)$ coincides with $\mbu(T-1)$ and $\beta(T-1; T) = 1$.  This case encompasses the single-stage ($T=1$) policy described in Section~\ref{subsec:policy1} and the last-stage allocation discussed at the beginning of Section~\ref{subsec:policyMulti}.  For $T > 1$, multipliers are reused across policies with different numbers of stages to reduce the number of degrees of freedom.  Specifically, %as in \cite{adapEstJSTSP2013},
\begin{equation}\label{eqn:beta}
\beta(t; T) = \beta(t-1; T-1), \quad t = 1, 2, \dots, T-2.
\end{equation}
The remaining first-stage multiplier $\beta(0; T)$ is optimized in a manner similar to \eqref{eqn:firstStage}.  Define $R^{\mathrm{OLFC-}T}(\mbxi(1))$ to be the Bayes risk cost of a $T$-stage OLFC allocation policy starting from stage $1$ and belief state $\mbxi(1)$ and using multipliers equal to those of a previously determined $T-1$-stage policy \eqref{eqn:beta}.  Then 
\begin{equation}\label{eqn:beta0}
\beta(0; T) = \argmin_{\beta\in[0,1]} \;\; \E\left[ R^{\mathrm{OLFC-}T}(\mbxi(1)) \mid \mbxi(0), \beta\mbv^{\ast}(0) \right].
\end{equation}
%The multipliers $\beta(t)$ are optimized 
This one-dimensional optimization can be carried out offline using Monte Carlo samples both to approximate the expectation as well as to simulate the cost $R^{\mathrm{OLFC}}(\mbxi(1))$ of the policy from stage $1$ onward. %as described in \cite{adapEstJSTSP2013}.  
As shown in \cite[Prop.~2]{adapEstJSTSP2013}, an important property of the procedure summarized by \eqref{eqn:beta}--\eqref{eqn:beta0} is that the resulting OLFC policies improve monotonically with the number of stages $T$.  Further details can be found in \cite{adapEstJSTSP2013}.

\section{Numerical Results}
\label{sec:num}

The multistage resource allocation policies described in Section~\ref{sec:policy} are numerically compared to the %recently proposed 
distilled sensing (DS) \cite{haupt2011} and sequential thresholding (ST) \cite{malloyarXiv2012} procedures, as well as to a single-stage non-adaptive baseline policy (NA).  For the results presented below, the number of hypothesis tests $n$ is $10^{4}$ and a homogeneous prior is used: $p_{i}(0) = p(0)$, $\mu_{i}^{0}(0) %= \mu^{0}(0) 
= 0$, $\mu_{i}^{1}(0) = 1$, $\sigma_{i}^{0}(0)^{2} = 0$, and $\sigma_{i}^{1}(0)^{2} = 1/16$ for all $i$.  Observations are simulated according to \eqref{eqn:obs} and \eqref{eqn:x}. The observation noise parameter $\nu^{2}$ is normalized to $1$ and the average budget per test $B$ is varied.  Since $\nu^{2}$ and %the resource allocations 
$u_{i}(t)$ always appear in the same ratio as in \eqref{eqn:obs}, an equivalent alternative would be to fix $B$ and vary $\nu^{2}$ instead.  The performance metric is %given by 
\eqref{eqn:BayesRisk} with $c = 1$, i.e., it is the expected number of errors of either type.

\begin{figure}[ht]
\centerline{
\subfigure[]{\includegraphics[width=0.40\columnwidth]{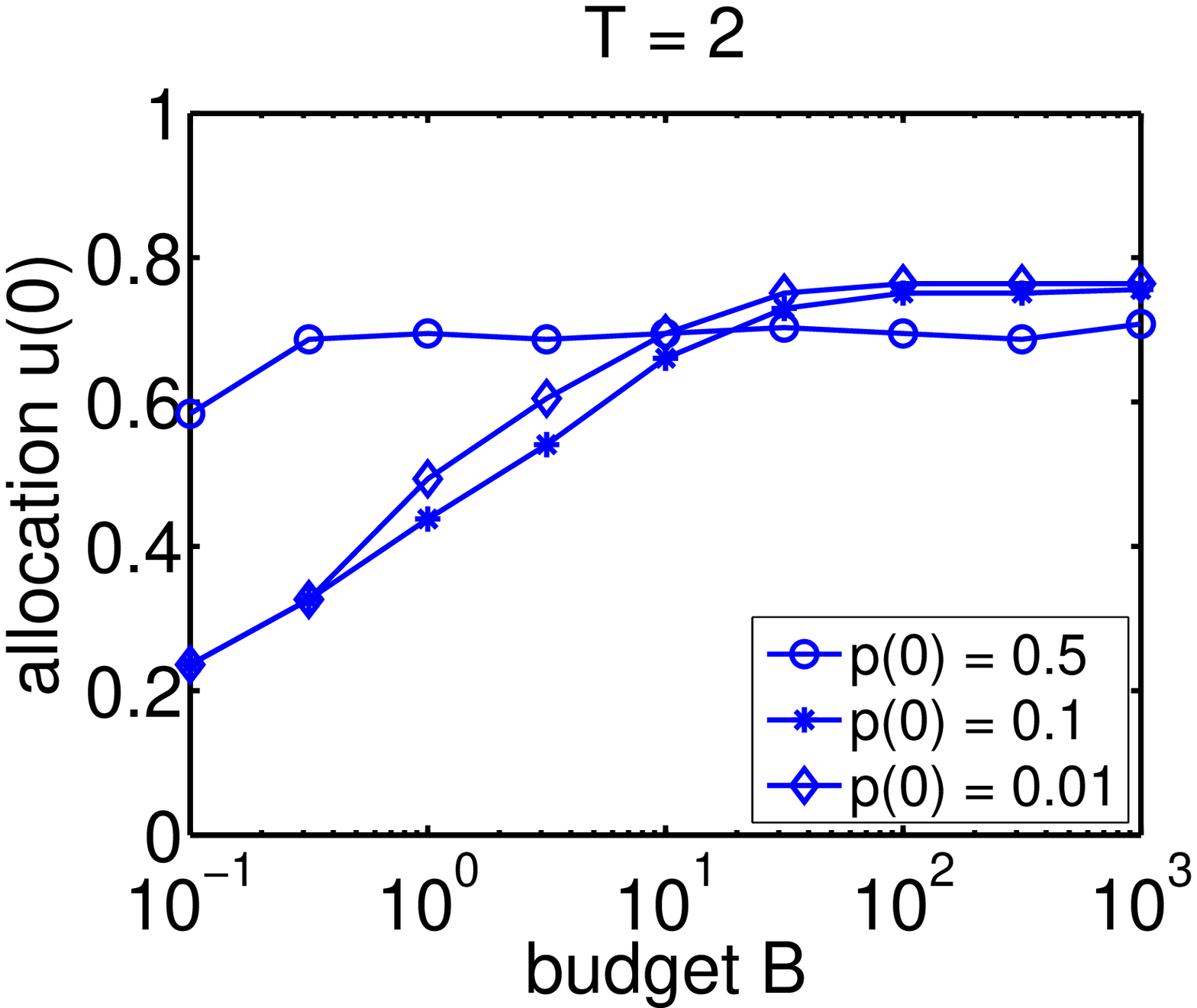}
\label{fig:beta2}}
\subfigure[]{\includegraphics[width=0.40\columnwidth]{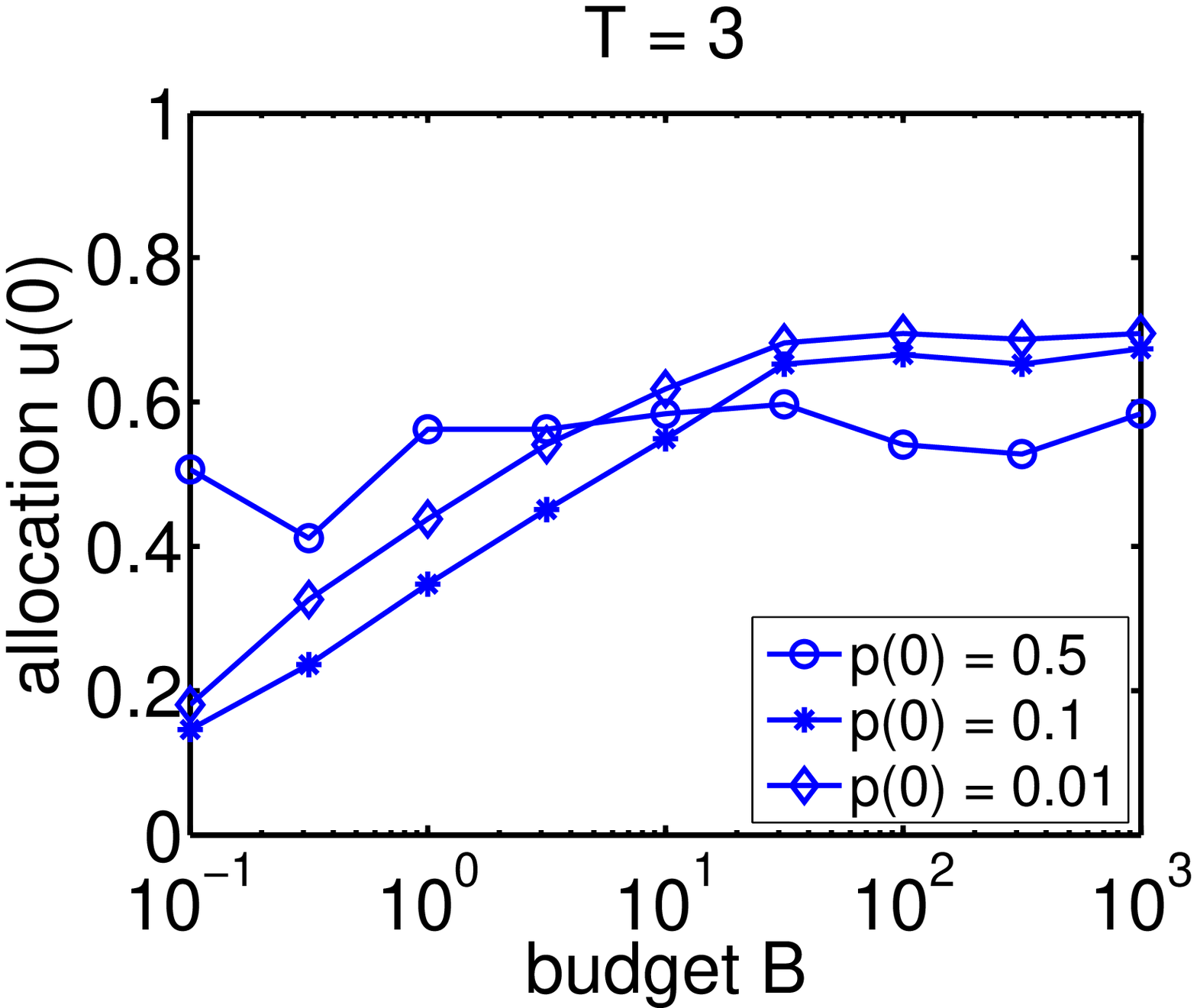}
\label{fig:beta3}}}
\caption{First-stage %resource 
allocation $u_{i}(0) = u(0)$ in the proposed $2$-stage (a) and $3$-stage (b) policies as a function of the mean proportion $p(0)$ of alternative hypotheses and the resource budget per test $B$.}
\label{fig:beta}
\end{figure}

The number of stages in the proposed OLFC policies is limited between 2 and 4.  In all cases, the first-stage allocation $\mbu(0)$ is uniform because of the homogeneous prior.  For $T = 2$, Fig.~\ref{fig:beta2} shows the first-stage budget fraction $u(0)$ that results from the offline optimization \eqref{eqn:firstStage} for different values of $p(0)$ and $B$.  Performance is not too sensitive to the exact value of $u(0)$ since the objective function in \eqref{eqn:firstStage} tends to be relatively flat away from the extremes $u(0) = 0$ and $u(0) = 1$.  Fig.~\ref{fig:beta3} plots the same parameter $u(0)$ for $T = 3$.  

For DS and ST, while \cite{haupt2011,malloyarXiv2012} prescribe values for $T$ as functions of $n$, in these experiments all $T \in \{2,\dots,12\}$ %values between $T = 2$ and $T = 12$ 
are tested and results for the best $T$ are shown. %presented.  
A similar optimization %search 
is performed over the %elimination 
parameter $\rho \in \{0.5, 0.6, 0.7, 0.8, 0.9\}$ in \cite{malloyarXiv2012}.  The budget allocations %allocations of the budget 
over stages follow \cite[eq.~(4),(5)]{haupt2011} and \cite[eq.~(14)]{malloyarXiv2012} respectively, except in the last stage of ST where the remaining budget is used up entirely.  
Two versions of DS and ST are implemented: the versions originally proposed in \cite{haupt2011,malloyarXiv2012} that use only the last stage of observations to make decisions, and Bayesian versions (DSB, STB), not proposed in \cite{haupt2011,malloyarXiv2012}, in which the %resource 
allocations $\mbu(t)$ are specified by \cite{haupt2011,malloyarXiv2012} but inference is done through the posterior update equations \eqref{eqn:postEvol}, thus incorporating all stages of observations.  As seen below, the Bayesian versions perform considerably better.

\begin{figure}[!ht]
\begin{comment}
\centerline{
\subfigure[]{\includegraphics[width=0.33\columnwidth]{Rs1r1.eps}
\label{fig:Rs1r1}}
\subfigure[]{\includegraphics[width=0.33\columnwidth]{Rs2r1.eps}
\label{fig:Rs2r1}}
\subfigure[]{\includegraphics[width=0.33\columnwidth]{Rs3r1.eps}
\label{fig:Rs3r1}}}
\centerline{
\subfigure[]{\includegraphics[width=0.33\columnwidth]{Rs1r2.eps}
\label{fig:Rs1r2}}
\subfigure[]{\includegraphics[width=0.33\columnwidth]{Rs2r2.eps}
\label{fig:Rs2r2}}
\subfigure[]{\includegraphics[width=0.33\columnwidth]{Rs3r2.eps}
\label{fig:Rs3r2}}}
\end{comment}
%\begin{comment}
\centerline{
\subfigure[]{\includegraphics[width=0.40\columnwidth]{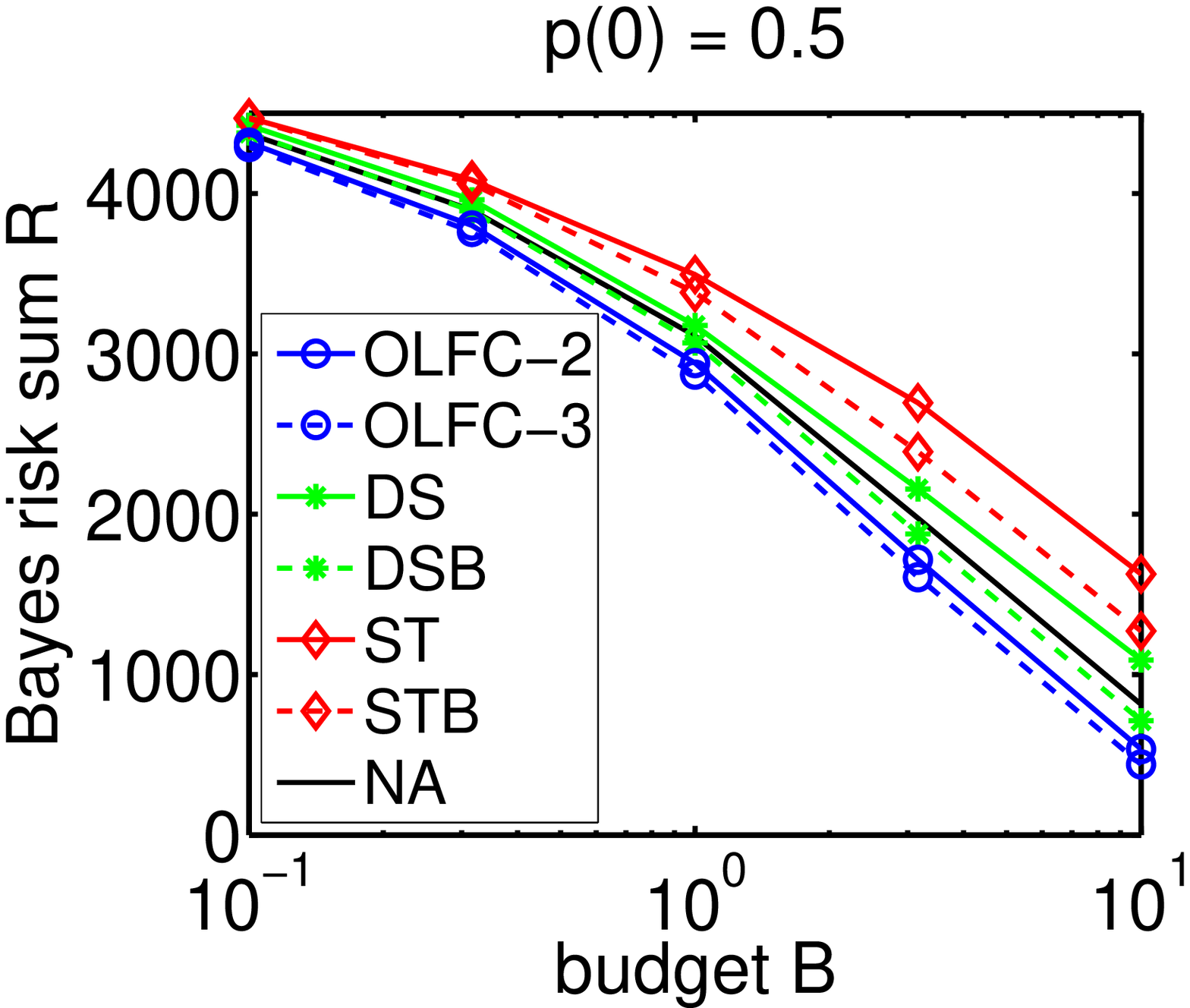}
\label{fig:Rs1r1}}
\subfigure[]{\includegraphics[width=0.40\columnwidth]{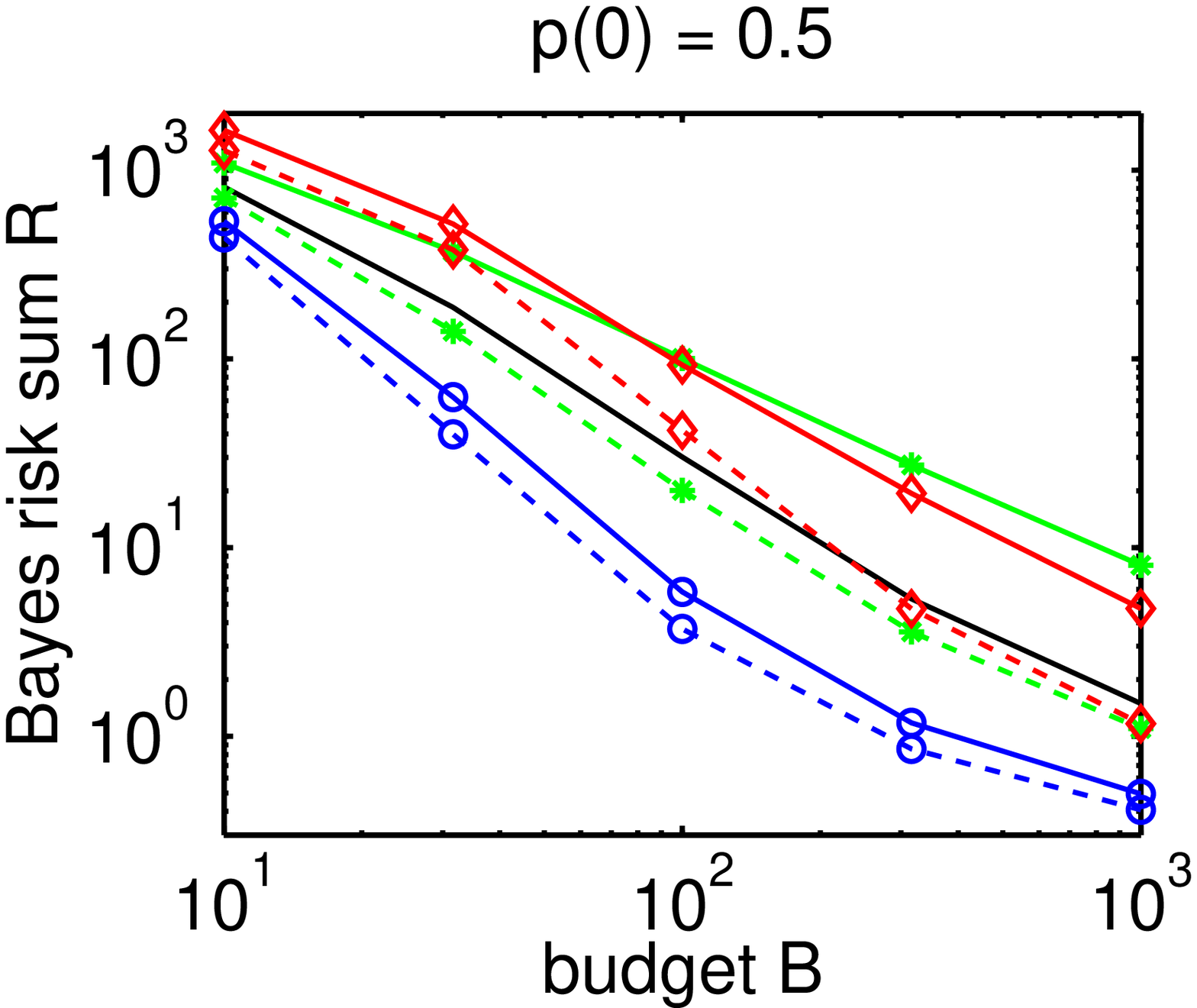}
\label{fig:Rs1r2}}}
\centerline{
\subfigure[]{\includegraphics[width=0.40\columnwidth]{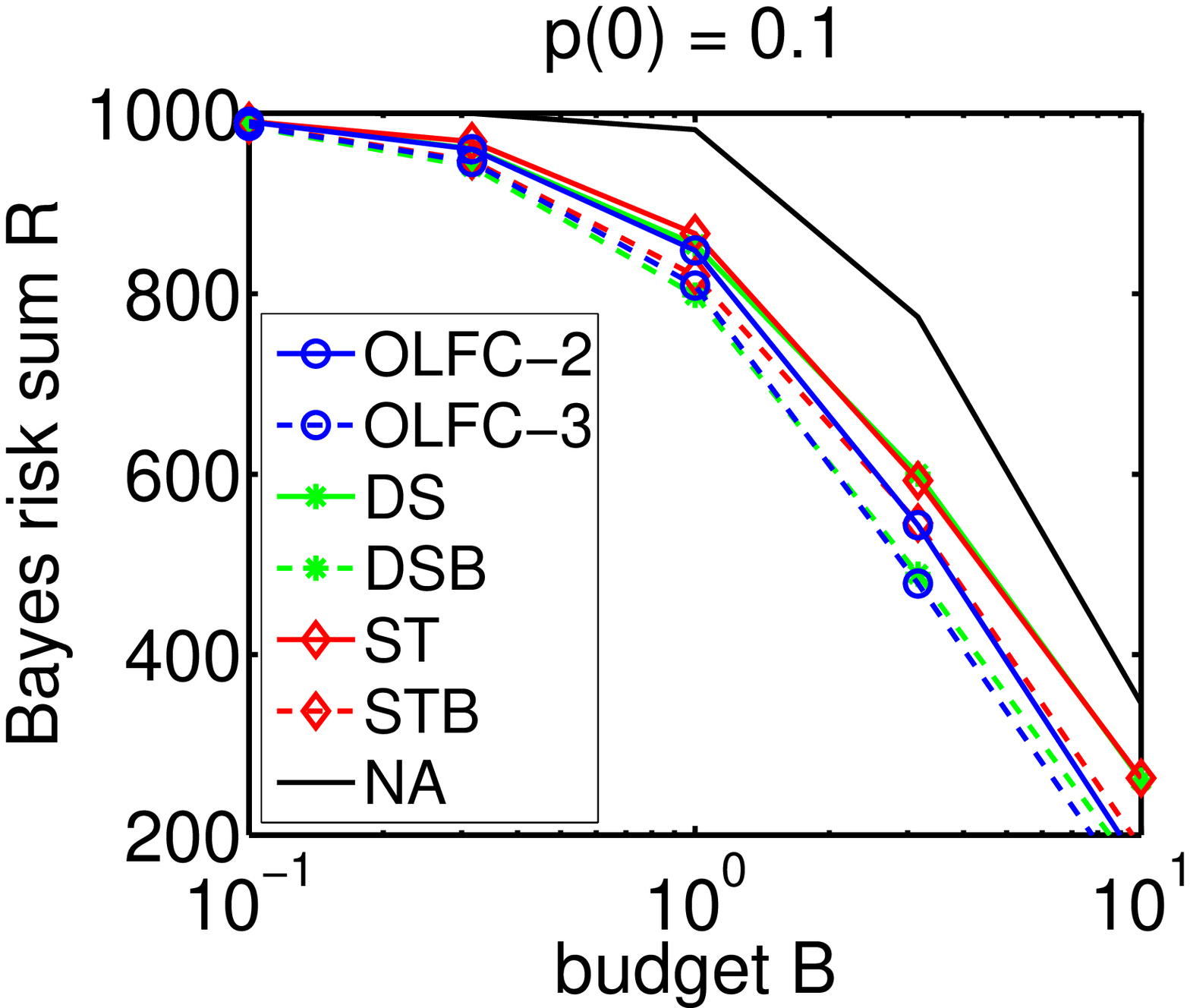}
\label{fig:Rs2r1}}
\subfigure[]{\includegraphics[width=0.40\columnwidth]{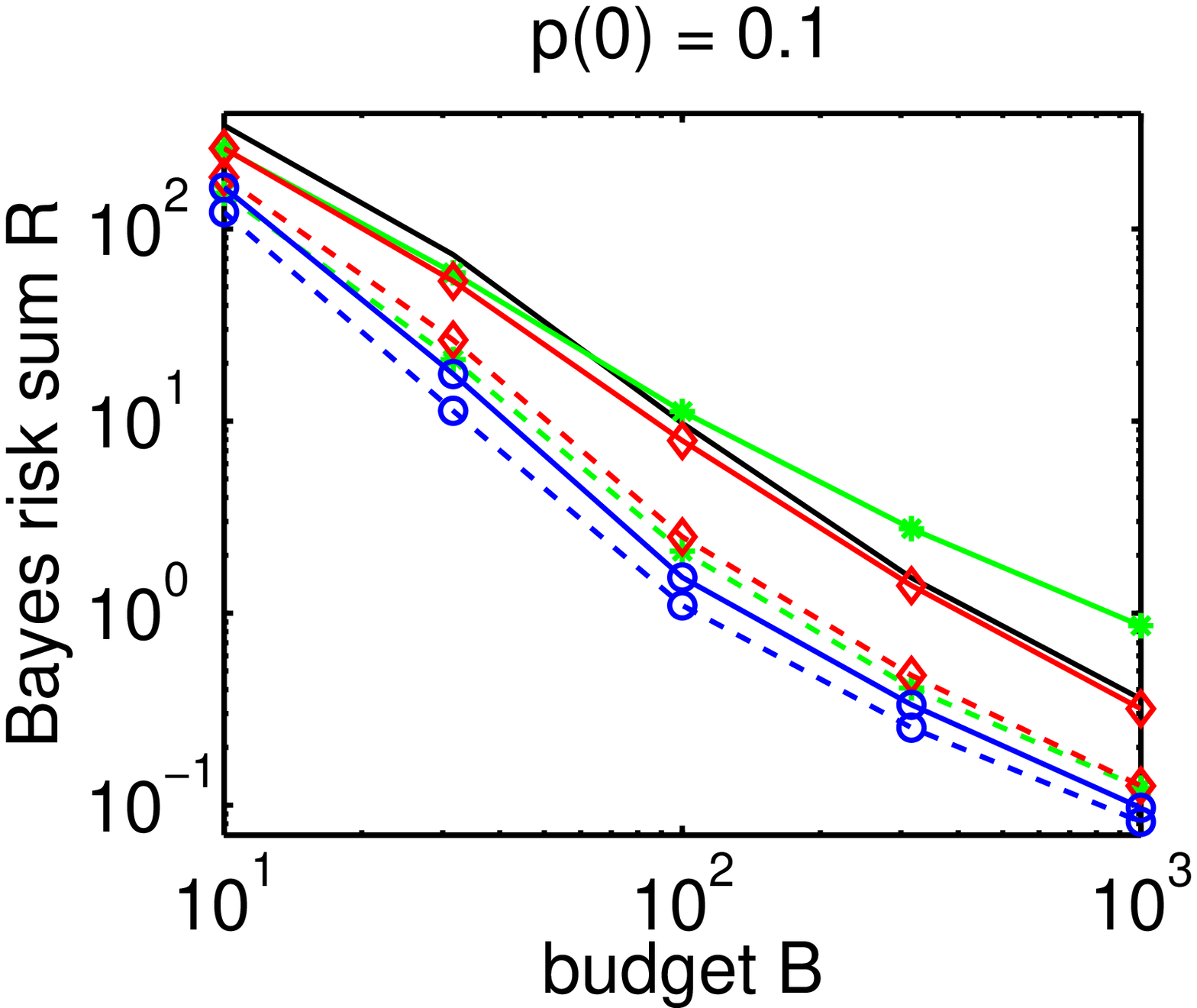}
\label{fig:Rs2r2}}}
\centerline{
\subfigure[]{\includegraphics[width=0.40\columnwidth]{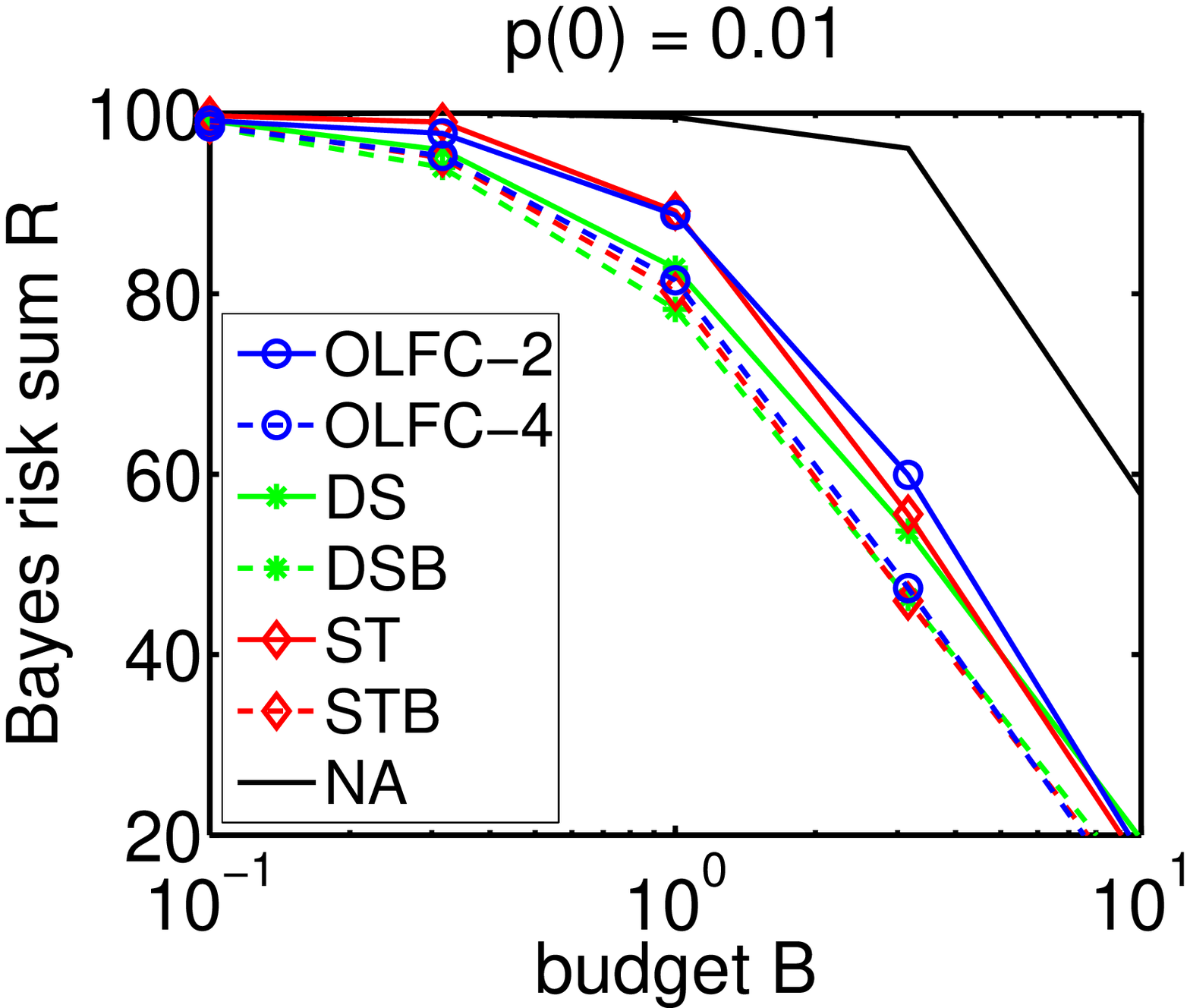}
\label{fig:Rs3r1}}
\subfigure[]{\includegraphics[width=0.40\columnwidth]{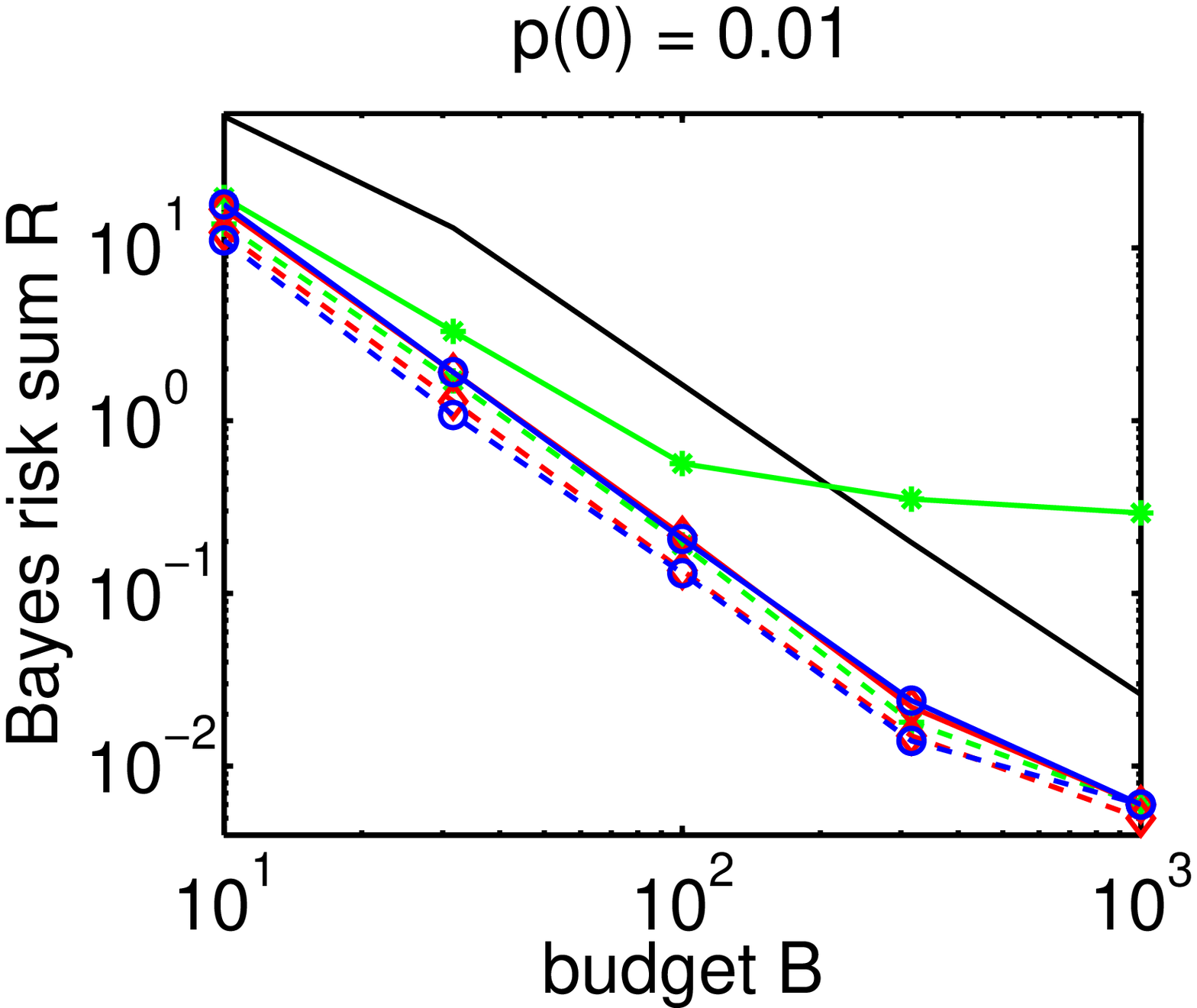}
\label{fig:Rs3r2}}}
%\end{comment}
\caption{Expected number of errors (\eqref{eqn:BayesRisk} with $c = 1$) resulting from the proposed open-loop feedback control policies with $T$ stages (OLFC-$T$), original and Bayesian versions of distilled sensing (DS, DSB) and sequential thresholding (ST, STB), and a non-adaptive baseline (NA).  The legends in (a), (c), (e) also apply to (b), (d), (f) respectively.  For $p(0) = 0.5, 0.1$ in (a)--(d), OLFC-$2$ and/or OLFC-$3$ outperform the alternative methods across budget levels. %DS, ST, and the non-adaptive policy  respectively.  
For $p(0) = 0.01$ in (e)(f), OLFC-$4$ is competitive with DSB and STB, while OLFC-$2$ %already 
achieves most of the gains using only $2$ stages.}
\label{fig:R}
%\vskip -2mm
\end{figure}

The performance of the policies is compared in Fig.~\ref{fig:R}.  For equiprobable hypotheses, $p(0) = 0.5$, the proposed $2$-stage policy achieves significant reductions in error (up to a factor of $5$) relative to the baseline NA policy, while the $3$-stage OLFC policy yields further improvement.  Since DS(B) and ST(B) are not designed for this non-sparse scenario, they perform less well, in some cases worse than NA.  For $p(0) = 0.1$, the $3$-stage OLFC policy essentially dominates the other policies, and at moderate to large resource levels in Fig.~\ref{fig:Rs2r2}, it is joined by the $2$-stage OLFC policy.  For $p(0) = 0.01$ and low resources in Fig.~\ref{fig:Rs3r1}, DSB and STB have slightly lower error rates than the $4$-stage OLFC policy, while for higher resources in Fig.~\ref{fig:Rs3r2}, the opposite is true.  Moreover, the $2$-stage OLFC policy attains most of the gains of these best-performing policies that use more stages.  In particular, the optimized DSB and STB policies shown in Fig.~\ref{fig:R} for $B \leq 1$ use at least $8$ and $6$ stages respectively. % in the low-resource regime 
%largest gains seem to occur at moderate resource levels

%\clearpage

\section{Conclusion}
\label{sec:concl}

This paper has explored the benefits of adaptive sensing for multiple binary hypothesis testing, notably in the regimes of balanced null and alternative hypotheses and few %a small number of 
allocation stages.  Future work includes generalizations to non-Gaussian observations, refinements of both the single-stage optimization and multistage dynamic programming procedures, and theoretical analysis that aims especially to understand the gains in the non-sparse setting and at moderate, non-asymptotic resource levels.

\appendices
\numberwithin{equation}{section}
\section{Bayes Risk Computation}
\label{app:BayesRisk}

This appendix derives the optimal Bayes risk for two Gaussian distributions, i.e., the integral in \eqref{eqn:BayesRiskDP} denoted as $R_{i}(u_{i}; \mbxi_{i})$ in Section~\ref{subsec:policy1}.  

To simplify notation in this appendix, both the stage index $T-1$ and test index $i$ are dropped.  Furthermore, we define $\mu \equiv \mu^{1} - \mu^{0}$ and shift the distributions so that $\mu^{0} = 0$ and $\mu^{1} = \mu$, without changing the Bayes risk.  Define $\Sigma^{H_{i}} = (\sigma^{H_{i}})^{2} + \nu^{2} / u$, $H_{i} = 0,1$, to be the conditional variances in \eqref{eqn:yPost}.  Recalling from Section~\ref{sec:prob} the assumption that $\sigma_{i}^{0}(0) \leq \sigma_{i}^{1}(0)$, it can be seen that $\Sigma^{0} \leq \Sigma^{1}$.  Two cases are considered.

{\bf Case $\Sigma^{0} < \Sigma^{1}$:}  First the decision regions 
\begin{align*}
\mcY^{0} &= \{ (1-p) f^{0}(y) \geq cp f^{1}(y) \},\\
\mcY^{1} &= \{ (1-p) f^{0}(y) < cp f^{1}(y) \}
\end{align*}
are determined, corresponding to the two terms in the minimization in \eqref{eqn:BayesRiskDP}.  Taking logarithms and collecting terms gives the following quadratic inequalities for $\mcY_{0}$:
\begin{gather}
\log(1-p) - \frac{1}{2} \log \Sigma^{0} - \frac{y^{2}}{2\Sigma^{0}} \geq \log(cp) - \frac{1}{2} \log \Sigma^{1} - \frac{(y - \mu)^{2}}{2\Sigma^{1}},\nonumber\\
\left( \frac{1}{2\Sigma^{0}} - \frac{1}{2\Sigma^{1}} \right) y^{2} + \frac{\mu}{\Sigma^{1}} y - \frac{\mu^{2}}{2\Sigma^{1}} - \frac{1}{2} \log \frac{\Sigma^{1}}{\Sigma^{0}} -\log\frac{1-p}{cp} \leq 0,\label{eqn:decisionRegionQuad}
\end{gather}
with the inequalities reversed for $\mcY_{1}$.  %Consider first the case $\Sigma^{0} < \Sigma^{1}$.  
Applying the quadratic formula to \eqref{eqn:decisionRegionQuad} yields the decision boundaries 
\begin{align*}
y_{\pm} &= \frac{\Sigma^{0} \Sigma^{1}}{\Sigma^{1} - \Sigma^{0}} \left( -\frac{\mu}{\Sigma^{1}} \pm \sqrt{ \left(\frac{\mu}{\Sigma^{1}}\right)^{2} + \frac{\Sigma^{1} - \Sigma^{0}}{\Sigma^{0} \Sigma^{1}} \left( \frac{\mu^{2}}{\Sigma^{1}} + \log \frac{\Sigma^{1}}{\Sigma^{0}} + 2\log\frac{1-p}{cp} \right) } \right)\\
&= \frac{-\Sigma^{0} \mu \pm \sqrt{\Sigma^{0} \Sigma^{1} D}}{\Sigma^{1} - \Sigma^{0}},
\end{align*}
provided that the discriminant
\[
D = \mu^{2} + \bigl(\Sigma^{1} - \Sigma^{0}\bigr) \left( \log \frac{\Sigma^{1}}{\Sigma^{0}} + 2\log\frac{1-p}{cp} \right)
\]
is non-negative.  The region $\mcY^{0}$ is the interval $[y_{-}, y_{+}]$ while the region $\mcY^{1}$ is the union of intervals $(-\infty, y_{-}) \cup (y_{+}, \infty)$.  If $D < 0$, then the decision boundaries do not exist, $\mcY^{0} = \emptyset$, and $\mcY^{1} = \mathbb{R}$.

Next the integrals of $f^{0}(y)$ and $f^{1}(y)$ are evaluated over $\mcY^{1}$ and $\mcY^{0}$ respectively, corresponding to the Type I and Type II error probabilities.  By standardizing the decision boundaries $y_{\pm}$, the Type I error probability can be expressed in terms of the standard Gaussian CDF $\Phi$ as 
\[
\mbbP^{0}(\mcY^{1}) = \Phi\left( -\frac{y_{+}}{\sqrt{\Sigma^{0}}} \right) + \Phi\left( \frac{y_{-}}{\sqrt{\Sigma^{0}}} \right) = \Phi\left( \frac{\sqrt{\Sigma^{0}} \mu - \sqrt{\Sigma^{1} D}}{\Sigma^{1} - \Sigma^{0}} \right) + \Phi\left( \frac{-\sqrt{\Sigma^{0}} \mu - \sqrt{\Sigma^{1} D}}{\Sigma^{1} - \Sigma^{0}} \right).
\]
Similarly the Type II error probability is 
\[
\mbbP^{1}(\mcY^{0}) = \Phi\left( \frac{y_{+} - \mu}{\sqrt{\Sigma^{1}}} \right) - \Phi\left( \frac{y_{-} - \mu}{\sqrt{\Sigma^{1}}} \right) = \Phi\left( \frac{-\sqrt{\Sigma^{1}} \mu + \sqrt{\Sigma^{0} D}}{\Sigma^{1} - \Sigma^{0}} \right) - \Phi\left( \frac{-\sqrt{\Sigma^{1}} \mu - \sqrt{\Sigma^{0} D}}{\Sigma^{1} - \Sigma^{0}} \right).
\]
The Bayes risk is then given by the linear combination 
\begin{equation}\label{eqn:BayesRiskLinCombo}
R_{i}(u; \mbxi) = (1-p) \mbbP^{0}(\mcY^{1}) + cp \mbbP^{1}(\mcY^{0}). 
\end{equation}

{\bf Case $\Sigma^{0} = \Sigma^{1} = \Sigma$:} In this case \eqref{eqn:decisionRegionQuad} simplifies to 
\[
y \leq \frac{\mu}{2} + \frac{\Sigma}{\mu} \log \frac{1-p}{cp} \equiv y_{c}
\]
for region $\mcY^{0}$, and $y > y_{c}$ for region $\mcY^{1}$.  The error probabilities are therefore 
\begin{align*}
\mbbP^{0}(\mcY^{1}) &= \Phi\left( -\frac{y_{c}}{\sqrt{\Sigma}} \right) = \Phi\left( -\frac{\mu}{2\sqrt{\Sigma}} - \frac{\sqrt{\Sigma}}{\mu} \log \frac{1-p}{cp} \right),\\
\mbbP^{1}(\mcY^{0}) &= \Phi\left( \frac{y_{c} - \mu}{\sqrt{\Sigma}} \right) = \Phi\left( -\frac{\mu}{2\sqrt{\Sigma}} + \frac{\sqrt{\Sigma}}{\mu} \log \frac{1-p}{cp} \right),
\end{align*}
and the Bayes risk is still given by \eqref{eqn:BayesRiskLinCombo}.

% you can choose not to have a title for an appendix
% if you want by leaving the argument blank
%\section{}
%Appendix two text goes here.

% use section* for acknowledgement
%\section*{Acknowledgment}

%The authors would like to thank...

% Can use something like this to put references on a page
% by themselves when using endfloat and the captionsoff option.
\ifCLASSOPTIONcaptionsoff
  \newpage
\fi

% trigger a \newpage just before the given reference
% number - used to balance the columns on the last page
% adjust value as needed - may need to be readjusted if
% the document is modified later
%\IEEEtriggeratref{8}
% The "triggered" command can be changed if desired:
%\IEEEtriggercmd{\enlargethispage{-5in}}

% references section

% can use a bibliography generated by BibTeX as a .bbl file
% BibTeX documentation can be easily obtained at:
% http://www.ctan.org/tex-archive/biblio/bibtex/contrib/doc/
% The IEEEtran BibTeX style support page is at:
% http://www.michaelshell.org/tex/ieeetran/bibtex/
\bibliographystyle{IEEEtran}
% argument is your BibTeX string definitions and bibliography database(s)
\bibliography{IEEEabrv,adapSens,optimization}
\end{document}